\documentclass[aip, apl, reprint]{revtex4-1}

\usepackage[utf8]{inputenc}
\usepackage[T1]{fontenc}
\usepackage[english]{babel}
\usepackage{graphicx}
\usepackage{epstopdf}
\usepackage{amsmath}
\usepackage{amssymb}
\usepackage{amstext}
\usepackage{amsfonts}
\usepackage{textcomp}
\usepackage[amssymb]{SIunits}
\usepackage{hyperref}
\usepackage{natbib}

\hypersetup{colorlinks=true, linkcolor=blue, citecolor=blue, urlcolor=blue}

\begin{document}

\title{Performance Boost in Industrial Multifilamentary Nb$_{\text{3}}$Sn Wires due to Radiation Induced Pinning Centers}

\author{T.\ Baumgartner}
\email{tbaumgartner@ati.ac.at}
\author{M.\ Eisterer}
\author{H.\ W.\ Weber}
\affiliation{Atominstitut, Vienna University of Technology, Stadionallee 2, 1020 Vienna, Austria}

\author{R.\ Fl{\"u}kiger}
\author{C.\ Scheuerlein}
\author{L.\ Bottura}
\affiliation{CERN, 1211 Geneva 23, Switzerland}

\date{\today}

\begin{abstract}

We report non-Cu critical current densities of $4.09 \cdot 10^9$A/m$^2$ at 12\,T and $2.27 \cdot 10^9$A/m$^2$ at 15\,T obtained from transport measurements on a Ti-alloyed RRP Nb$_3$Sn wire after irradiation to a fast neutron fluence of $8.9 \cdot 10^{21}$\,m$^{-2}$. These values are to our knowledge unprecedented in multifilamentary Nb$_3$Sn, and they correspond to a $J_{\text{c}}$ enhancement of approximately 60\% relative to the unirradiated state. Our magnetometry data obtained on short wire samples irradiated to fast neutron fluences of up to $2.5 \cdot 10^{22}$\,m$^{-2}$ indicate the possibility of an even better performance, whereas earlier irradiation studies on bronze-processed Nb$_3$Sn wires with a Sn content further from stoichiometry attested a decline of the critical current density at such high fluences. We show that radiation induced point-pinning centers rather than an increase of the upper critical field are responsible for this $J_{\text{c}}$ enhancement, and argue that these results call for further research on pinning landscape engineering.

\end{abstract}


\maketitle


\section*{Introduction}

Multifilamentary Nb$_3$Sn wires are currently the most appropriate superconductors for applications requiring magnetic fields beyond the capabilities of Nb-Ti wires (${\gtrsim}10$\,T). Through extensive research efforts in the fields of materials science and superconductor technology they have come a long way from the early bronze-processed strands to state-of-the-art restack rod processed (RRP) and powder-in-tube (PIT) wires. The former were limited to non-matrix critical current densities below $10^9$\,A/m$^2$ at 4.2\,K and 12\,T, whereas the latter achieve $3 \cdot 10^9$\,A/m$^2$ at the same temperature and field, while having superior high-field properties due to upper critical field optimization by ternary element addition. \cite{Dietderich:Nb3Sn_research, Godeke:PIT}

So far the two main strategies employed for improving the critical current density $J_{\text{c}}$ were grain refinement and impurity doping. The former strives to decrease the grain size by optimizing the heat treatment process, resulting in a higher density of grain boundaries, which have been known to be the dominant pinning sites in Nb$_3$Sn for a long time. \cite{Scanlan:flux_pinning_centers} The latter is aimed at improving the upper critical field $B_{\text{c2}}$ by introducing additives -- usually tantalum or titanium -- which increase the normal-state resistivity. \cite{Fluekiger:microstructure} The upper critical field optimization is already very efficient in modern wires, although the $B_{\text{c2}}$ homogeneity can probably still be improved by minimizing Sn concentration gradients. Reducing the grain size further by lowering the heat treatment temperature would compromise the homogeneity of the \mbox{A-15} layer, resulting in an adverse effect on $J_{\text{c}}$. \cite{Tarantini:Nb3Sn_optimization} Recently Xu et al.\ employed an alternative grain refinement strategy based on the introduction of ZrO$_2$ nano-particles. Their results are promising, however, the very high critical current densities they attained were limited to a thin layer of fine grains, which constitutes only a small fraction of the strand cross section. \cite{Xu:ZrO2_in_Nb3Sn, Xu:internally_oxidized}

In this paper we present experimental data suggesting that the introduction of point-pinning centers, i.e.\ defects which are small compared to the inter-vortex spacing, is a viable route for further increasing the critical current density of Nb$_3$Sn, especially at high magnetic fields. These data were obtained on state-of-the-art commercial multifilamentary wires subjected to fast neutron irradiation ($E > 0.1$\,MeV), which is known to create disordered regions a few nanometers in size. \cite{Pande:Tc_irradiation, Kuepfer:summation, Holdway:radiation_damage} We found that in sufficient quantity these defects have a pronounced impact on both the magnitude and the functional dependence of the volume pinning force in the examined wires, causing a significant increase of the high-field $J_{\text{c}}$. These results suggest that the limits of Nb$_3$Sn wires can be pushed further by engineering their pinning landscape, for instance by the introduction of nano-particles.

Five types of multifilamentary Nb$_3$Sn wires, differing in terms of production process and additive element, were included in our neutron irradiation study. We will, however, focus on one of them, since it is the only sample type on which transport critical current measurements were performed after irradiation (the others were characterized by means of SQUID magnetometry). This wire is a Ti-alloyed RRP strand produced by Oxford Superconducting Technology, and was reacted using an optimized heat treatment (210\,{\textcelsius} for 48\,h, 400\,{\textcelsius} for 48\,h, 665\,{\textcelsius} for 50\,h). It has a diameter of 0.817\,mm (after the heat treatment), and contains 108 sub-elements embedded in a copper matrix in four shells (stacking configuration 18--24--30--36). The Cu/non-Cu volume ratio of the wire is 1.02, corresponding to a non-Cu cross section of 0.260\,mm$^2$.


\section*{Results}

\begin{figure}
	\includegraphics[scale=1.]{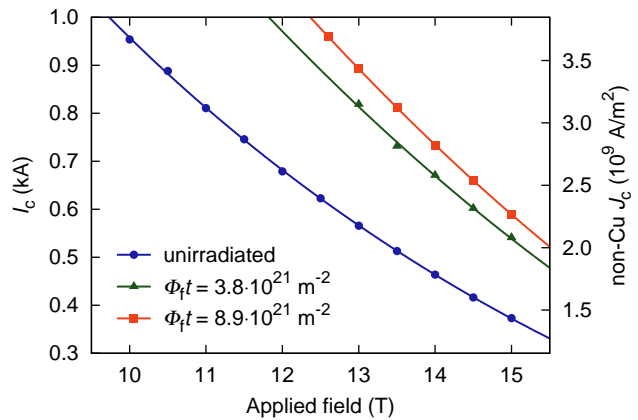}%
	\caption{Transport critical current and corresponding non-Cu critical current density as a function of applied field, obtained before irradiation and at the two specified fast neutron fluences. The points represent the actual experimental data, the solid lines are fits.
	\label{fig:Ic_trans_irrad}}
\end{figure}

Transport critical current data obtained on the examined wire in the unirradiated state as well as after irradiation to fast neutron fluences of $\varPhi_{\text{f}} t = 3.8 \cdot 10^{21}$\,m$^{-2}$ and $\varPhi_{\text{f}} t = 8.9 \cdot 10^{21}$\,m$^{-2}$ are shown in \autoref{fig:Ic_trans_irrad}. The data set of the unirradiated wire was compiled by averaging over results from six samples, including those which were irradiated later on, and has a standard deviation of only 1\%. The solid lines are fits which were computed using the well-known expression for the volume pinning force
\begin{equation}
	F_{\text{p}} = \vert \vec{B} \times \vec{J_{\text{c}}} \vert = C \, f(b) \:\:\: \text{with} \:\:\:  f(b) = b^p \, (1 - b)^q \; \text{,}
	\label{eq:Fp_USL}
\end{equation}
where $C$ is proportional to the maximum pinning force at the temperature of interest, and $f(b)$, often referred to as the pinning force function, describes the dependence on the reduced magnetic field $b = B / B_{\text{c2}}^*$, with $B_{\text{c2}}^* \approx B_{\text{c2}}$ in the case of Nb$_3$Sn. \cite{Ekin:USL} The scaling exponents $p$ and $q$ determine the shape of the pinning force function, and have been calculated analytically for different pinning mechanisms. \cite{Dew-Hughes:flux_pinning} For grain boundary pinning, which is generally assumed to be the dominant pinning mechanism in Nb$_3$Sn, the predicted exponents are $p = 0.5$ and $q = 2$.

In literature it is common to specify the applied field $B_{\text{a}}$ in the context of wire performance rather than the actual, self-field corrected value $B$ inside the wire, and we adhere to this convention in the present work. Since the self-field amounts to less than 5\% of the applied field in the field range shown in \autoref{fig:Ic_trans_irrad}, the error introduced by assuming $B = B_{\text{a}}$ when applying \autoref{eq:Fp_USL} is acceptable. While the values of $B_{\text{c2}}^*$, $p$, and $q$ obtained from fits to relatively small data sets cannot be expected to be very accurate, the fit functions shown in \autoref{fig:Ic_trans_irrad} are clearly suitable for extrapolating to values somewhat outside the field range of the measurements.

An examination of the results shown in \autoref{fig:Ic_trans_irrad} reveals three important features. First, with a non-Cu $J_{\text{c}}$ of $2.61 \cdot 10^9$A/m$^2$ at 12\,T and $1.44 \cdot 10^9$A/m$^2$ at 15\,T the wire already exhibits a high critical current density in the unirradiated state. Second, the sample subjected to the higher fluence reached $4.09 \cdot 10^9$A/m$^2$ at 12\,T (extrapolated using the fit function), and $2.27 \cdot 10^9$A/m$^2$ at 15\,T, corresponding to a significant increase of 56\% and 58\%, respectively. And third, the increase in $J_{\text{c}}$ between the lower and the higher fluence demonstrates that the fluence at which the maximum increase occurs is greater than $\varPhi_{\text{f}} t = 3.8 \cdot 10^{21}$\,m$^{-2}$. This suggests the potential for a $J_{\text{c}}$ enhancement at unexpectedly high fluences, which is discussed in the following.

\begin{figure}
	\includegraphics[scale=1.]{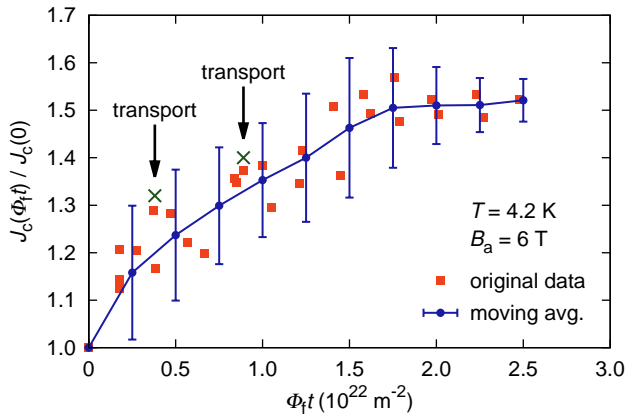}%
	\caption{Relative change of $J_{\text{c}}(4.2\,\text{K}, 6\,\text{T})$ as a function of fast neutron fluence, assessed by means of SQUID magnetometry. The squares represent the experimental data, the solid line is a moving average with error bars indicating $\pm 3 \sigma$.
	\label{fig:flu_Jc}}
\end{figure}

The transport results obtained from the sample irradiated to a fast neutron fluence of $8.9 \cdot 10^{21}$\,m$^{-2}$ are impressive, however, our magnetometry data indicate an even larger $J_{\text{c}}$ increase at higher fluences. In an earlier publication we demonstrated that the critical current density of a multifilamentary Nb$_3$Sn wire can be reliably evaluated from magnetometry data, if the sub-element geometry is known. \cite{Baumgartner:Jc_mag} This method was used to assess the fluence dependent change $J_{\text{c}}(\varPhi_{\text{f}} t) / J_{\text{c}}(0)$ of the critical current density relative to the unirradiated state using short wire samples. Note that differences in the absolute values of $J_{\text{c}}$ obtained from magnetometry and from transport measurements are irrelevant for evaluating the relative change exhibited by the same sample. Since the applied field of our SQUID magnetometer is limited to 7\,T, the evaluation was carried out for data obtained at 4.2\,K and 6\,T. The results are shown in \autoref{fig:flu_Jc} up to the highest fluence available at the time of writing, i.e.\ $2.5 \cdot 10^{22}$\,m$^{-2}$. The squares represent the actual $J_{\text{c}}(\varPhi_{\text{f}} t) / J_{\text{c}}(0)$ data derived from magnetometry, and the solid line is a moving average computed as described in the Methods section.

The two data points indicated by arrows show the $J_{\text{c}}$ enhancement at 6\,T extrapolated from our transport measurements. They were calculated using the fit functions shown in \autoref{fig:Ic_trans_irrad}, which can of course not be expected to be very accurate so far outside of the experimental field range. Judging from \autoref{fig:flu_Jc}, the maximum $J_{\text{c}}$ enhancement at 6\,T should occur somewhat above $2.5 \cdot 10^{22}$\,m$^{-2}$. Since this value is significantly higher than the fluences at which transport data are available, it stands to reason that an even higher enhancement than the approximately 60\% found in transport measurements in the 12\,--15\,T range can be attained. At $\varPhi_{\text{f}} t = 8.9 \cdot 10^{21}$\,m$^{-2}$, the higher one of the two transport sample fluences, the $J_{\text{c}}$ enhancement at 6\,T is roughly 35\%, whereas the peak value appears to be about 55\%. Transferring this difference to the $J_{\text{c}}$ enhancement at 12\,--15\,T yields a potential increase of 80\% at $\varPhi_{\text{f}} t = 2.5 \cdot 10^{22}$\,m$^{-2}$ relative to the unirradiated state, which translates to $J_{\text{c}}(4.2\,\text{K}, 12\,\text{T}) \approx 4.7 \cdot 10^9$\,A/m$^2$, and $J_{\text{c}}(4.2\,\text{K}, 15\,\text{T}) \approx 2.6 \cdot 10^9$\,A/m$^2$.

In earlier irradiation studies on Ti-alloyed Nb$_3$Sn wires produced by the bronze process the decrease of $J_{\text{c}}$ after reaching the maximum enhancement was found to occur at lower fluence values than the peak fluence of ${\sim}2.5 \cdot 10^{22}$\,m$^{-2}$ reported here. \cite{Weiss:Nb3Sn_irrad, Hahn:fusion_neutron_irradiation} Note that 14\,MeV neutron radiation was used in the cited works, which means that damage energy scaling must be employed in order to compare these results to ours. \cite{Weber:radiation_effects} The appropriate scaling factor to convert to the equivalent damage of neutrons with $E > 0.1$\,MeV is approximately 3.4, which puts the peak fluences of the earlier studies in the range $\varPhi_{\text{f}} t = 0.7 \text{--} 2.4 \cdot 10^{22}$\,m$^{-2}$. Based on the results of the cited studies and our own, we believe that the difference in peak fluence can be explained based on the pronounced influence of Ti additions on the fluence dependence of the $J_{\text{c}}$ enhancement and on the fact that the Sn content of the wires we studied is closer to stoichiometry.

\begin{figure}
	\includegraphics[scale=1.]{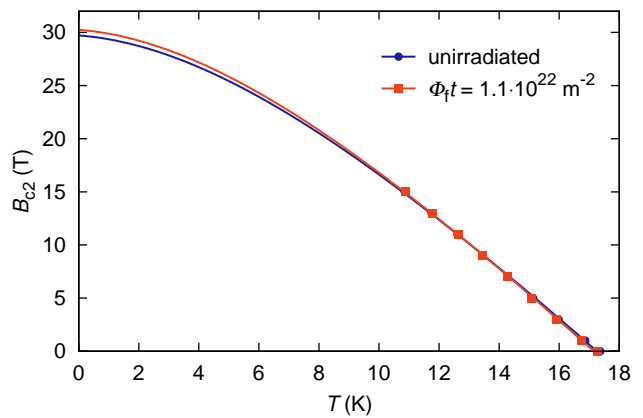}%
	\caption{Increase of the upper critical field of the examined wire due to fast neutron irradiation. Experimental data are displayed as points, and the solid lines are fits used to extrapolate to lower temperatures.
	\label{fig:Bc2_irrad}}
\end{figure}

The majority of earlier irradiation studies on Nb$_3$Sn attributed the radiation induced $J_{\text{c}}$ enhancement to an increase of the upper critical field, arguing that the disorder caused by irradiation increases the normal-state resistivity and with it $B_{\text{c2}}$, as predicted by the Gor'kov-Goodman relation. \autoref{fig:Bc2_irrad} shows the temperature dependence of the upper critical field of the examined wire in the unirradiated state and after irradiation to a fluence of $1.1 \cdot 10^{22}$\,m$^{-2}$. The points represent the experimental data, whereas the solid lines are fits based on the dirty limit $B_{\text{c2}}(T)$ dependence derived by Helfand and Werthamer. \cite{WHH:Hc2_II} As evident from the plot, the Nb$_3$Sn wire discussed in the present work exhibits only a slight $B_{\text{c2}}$ increase (2\% at $T = 4.2$\,K, obtained from the fits) after irradiation to a relatively high fluence.

Obviously this minor change in $B_{\text{c2}}$ cannot explain the large $J_{\text{c}}$ enhancement discussed above. The smallness of the increase suggests that the defect structure produced by fast neutron irradiation has little impact on the normal-state resistivity. This is consistent with the small decrease in the critical temperature (approximately 2\% at $\varPhi_{\text{f}} t = 10^{22}$\,m$^{-2}$), which correlates with the normal-state resistivity via electron lifetime effects. \cite{Testardi:electron_lifetime} A few earlier neutron irradiation studies suggested changes in the pinning behavior as the cause of the radiation induced $J_{\text{c}}$ increase. \cite{Cullen:neutron_irrad, Guinan:low_temp_irradiation, Okada:irrad} To our knowledge Cullen and Novak (cf.\ Ref.\ \onlinecite{Cullen:neutron_irrad}) were the first to claim that the radiation induced changes in the normal-state resistivity are insufficient to explain the observed $J_{\text{c}}$ enhancement. They concluded that neutron irradiation produces defects which act as pinning centers.

\section*{Discussion}

A detailed analysis of neutron radiation induced changes in the volume pinning force of the wire discussed in this work as well as the four others included in our irradiation study is given in an earlier publication. \cite{Baumgartner:irrad_scaling} We found that with increasing fluence the pinning force function shifts from the behavior expected for grain boundary pinning ($p = 0.5$, $q = 2$, cf.\ \autoref{eq:Fp_USL}) towards that expected for pinning by point-like defects ($p = 1$, $q = 2$). The pinning force function of irradiated samples was found to be in good agreement with the two-component ansatz
\begin{equation}
f(b) = \alpha \, b^{p_1} \, (1 - b)^{q_1} + \beta \, b^{p_2} \, (1 - b)^{q_2} \; \text{,}
\label{eq:f_2M}
\end{equation}
with the exponents $p_1$ and $q_1$ describing the unirradiated state, and the second term with the fixed exponents $p_2 = 1$ and $q_2 = 2$ modeling the radiation induced change in the pinning behavior. The relative weights $\alpha$ and $\beta$ indicate the respective contributions of the original pinning mechanism and of the radiation induced point-pinning. Within the examined fluence range a monotonous increase of the point-pinning parameter $\beta$ was observed.

\begin{figure}
	\includegraphics[scale=1.]{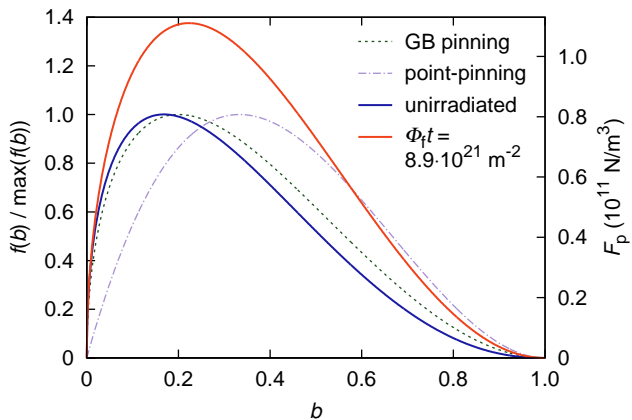}%
	\caption{Illustration of radiation induced changes in the volume pinning force. The dashed curves represent the normalized functional dependence predicted for grain boundary (GB) and for point-pinning (left scale). The solid lines show the actual volume pinning force in the unirradiated state and at the specified fluence, obtained from pinning force analyses of magnetometry data (right scale).
		\label{fig:Fp_shift}}
\end{figure}

The pinning force functions for grain boundary pinning and for point-pinning, both normalized to a peak value of unity, are shown as dashed and dot-dashed lines in \autoref{fig:Fp_shift}. Above a reduced field of $b \approx 0.27$ point-pinning is superior to grain boundary pinning, assuming identical peak values. In reality the pinning force function in the unirradiated state (solid blue line) exhibits certain deviations from the theoretical prediction, and the volume pinning force after irradiation (solid red line) does not only shift to the right, but also increases in magnitude, since the radiation induced pinning centers are added to the original ones instead of replacing them. The latter two curves were computed based on pinning force analyses of magnetometry data obtained in the temperature range from 4.2 to 15\,K, using the two-component pinning force model described in Ref.\ \onlinecite{Baumgartner:irrad_scaling}. From \autoref{fig:Fp_shift} it is obvious that a radiation induced point-pinning contribution and the according shift of the pinning force maximum towards higher field values can easily explain the observed $J_{\text{c}}$ enhancement.

Our transport critical current measurements directly show that non-Cu $J_{\text{c}}$ values of $4.09 \cdot 10^9$A/m$^2$ at 12\,T and $2.27 \cdot 10^9$A/m$^2$ at 15\,T are possible in Nb$_3$Sn wires containing radiation induced defects. To our knowledge such a performance of industrially produced multifilamentary Nb$_3$Sn is unprecedented in literature, and the data obtained from our extensive magnetometry study on sequentially irradiated short wire samples strongly indicate that even higher critical current densities are attainable. In fact, recent results obtained from a proton irradiation study on the same Ti-alloyed RRP wire show a $J_{\text{c}}$ enhancement of 100\% relative to the unirradiated state at an applied field of 10\,T. \cite{Spina:proton_irrad}

The essential change induced by fast neutron radiation is the addition of point-pinning centers, leading to an increase of the maximum volume pinning force and to a shift of the pinning force peak towards higher reduced field values. This point-pinning contribution results in a significant enhancement of $J_{\text{c}}$ over a wide field range, as shown in \autoref{fig:Fp_shift}. Considering the radioactivity arising from neutron irradiation, this technique can hardly be deemed suitable for optimizing commercial wires.

We believe, however, that the nature of the additional point-pinning centers is secondary, and that the same pinning optimization can be achieved in a different manner. A process which affects the functional dependence of the volume pinning force in a similar way to that reported here has recently been introduced by Xu et al., who achieved the precipitation of nanoscopic ZrO$_2$ particles in a Nb$_3$Sn strand using a Nb-Zr alloy and SnO$_2$ powder as precursor materials. \cite{Xu:ZrO2_in_Nb3Sn, Xu:internally_oxidized} Their promising results and our observation that defects a few nanometers in size produce a point-pinning contribution which adds to the grain boundary pinning, make us confident that after decades of research Nb$_3$Sn still has a considerable potential for improvement.


\section*{Methods} \label{sec:methods}

The neutron irradiation was carried out in the central irradiation facility of the TRIGA \mbox{Mark-II} reactor at Atominstitut. Small pieces of nickel foil were irradiated together with the samples for assessing the fast neutron fluence by means of gamma spectroscopy, making use of the $^{58}$Ni(n,p)$^{58}$Co reaction, which has a threshold energy of approximately 1\,MeV. Based on the known neutron spectrum of the reactor, this method allows the calculation of the fast neutron fluence, which we define in conformity with many other irradiation studies as the fluence of all neutrons with kinetic energies $E > 0.1$\,MeV. The short wire samples used for magnetometry were irradiated sequentially in fluence steps of about $2 \cdot 10^{21}$\,m$^{-2}$, whereas each transport sample was irradiated only once.

For the preparation of transport samples unreacted wire pieces about 0.5\,m in length were wound on mini-VAMAS barrels (23\,mm in diameter, 34\,mm in length) composed of the alloy Ti-6Al-4V. After carrying out the heat treatment, the last winding on either side was soldered to copper rings attached to the barrel. Transport critical current measurements were carried out in liquid helium at ambient pressure in applied fields of up to 15\,T. A sample rod equipped with current leads made of a high-temperature superconductor was used in combination with a 1000\,A current source to apply current to the samples. Low-resistance pressure contacts between the sample rod and the copper rings on the barrel were established using indium rings as interface material. An electric field criterion of 10\,{\micro}V/m was used to evaluate the critical current, which is approximately one order of magnitude lower than the current shut-down criterion used during the measurements.

For magnetization measurements straight pieces of wire were heat treated, and cut into ${\sim}$4\,mm long pieces using a low-speed diamond saw. These short wire samples were characterized in a Quantum Design MPMS XL SQUID magnetometer (7\,T maximum applied field) equipped with RSO (Reciprocating Sample Option) in the unirradiated state and after each irradiation step. Magnetization loops, which allow an evaluation of the critical current density from the irreversible magnetic moment, were recorded at temperatures ranging from 4.2 to 15\,K.

To cope with the relatively large scatter in the $J_{\text{c}}(\varPhi_{\text{f}} t) / J_{\text{c}}(0)$ data obtained from magnetometry samples, a moving average was obtained by calculating piecewise linear fits at equidistant sampling positions. For this purpose the data were convoluted with the Gaussian weighting function
\begin{equation}
w(\varPhi_{\text{f}} t) = \exp{\left( \frac{-(\varPhi_{\text{f}} t - \varPhi_{\text{f}} t_{\text{s}})^2}{2 (\varPhi_{\text{f}} \tau)^2} \right)} \; \text{,}
\label{eq:weighting}
\end{equation}
where $\varPhi_{\text{f}} t_{\text{s}}$ denotes the sampling position, and $\varPhi_{\text{f}} \tau$ is the characteristic width of the weighting function, which was set to $3 \cdot 10^{21}$\,m$^{-2}$. The values of the moving average function were calculated at the points along the curve marked with dots in \autoref{fig:flu_Jc} by evaluating the linear fits to the weighted data at these positions. The error bars in the graph indicate $\pm 3 \sigma$, where $\sigma$ is the standard deviation of the data relative to the linear fits, calculated with the same weights used for fitting.

The temperature dependence of the upper critical field was obtained from short wire samples by slowly ramping the temperature at various applied field values of up to 15\,T, and measuring the voltage drop across the sample caused by a current of 100\,mA. A transition mid-point criterion was used to evaluate the critical temperature $T_{\text{c}}$ at each field, which directly yields $B_{\text{c2}}(T)$. The extrapolations to low temperatures were computed using the function
\begin{equation}
	h_{\text{fit}}^*(t) = 1 - t - C_1 \, (1 - t)^2 - C_2 \, (1 - t)^4 \; \text{,}
	\label{eq:WHH_fit}
\end{equation}
with $C_1 = 0.153$ and $C_2 = 0.152$, which was found by the authors to deviate by less than 1\% from the dirty limit dependence of the upper critical field on the reduced temperature $t = T / T_{\text{c}}$ calculated by Helfand and Werthamer. \cite{WHH:Hc2_II} Using \autoref{eq:WHH_fit}, the upper critical field at arbitrary temperatures below $T_{\text{c}}$ is given by
\begin{equation}
	B_{\text{c2}}(T) = \frac{B_{\text{c2}}(0)}{0.693} \, h_{\text{fit}}^*(T / T_{\text{c}}) \; \text{.}
	\label{eq:Bc2_fit}
\end{equation}
The zero-temperature upper critical field $B_{\text{c2}}(0)$ and the critical temperature $T_{\text{c}}$ were used as fit parameters to obtain the curves shown in \autoref{fig:Bc2_irrad}.


\section*{Author Contributions}

L.B.\ and H.W.W.\ initiated this joint research project. C.S.\ was responsible for the sample preparation. M.E.\ supervised the activities at Atominstitut, where T.B.\ performed the measurements, wrote the main manuscript text, and prepared the figures. R.F.'s comprehensive knowledge of materials science and earlier irradiation studies on Nb$_3$Sn was crucial to this work. All authors discussed the results and commented on the manuscript.

\section*{Additional Information}

\subsection*{Competing Financial Interests}

The authors declare no competing financial interests.

\bibliography{Bibliography}

\end{document}